\documentclass{article}

\usepackage[nonatbib,final]{neurips_2019}

\usepackage[utf8]{inputenc} 
\usepackage[T1]{fontenc}    
\usepackage[hidelinks]{hyperref}       
\usepackage{url}            
\usepackage{booktabs}       
\usepackage{amsfonts}       
\usepackage{amssymb}
\usepackage{amsmath}
\usepackage{nicefrac}       
\usepackage{microtype}      
\usepackage{multirow}
\usepackage{xcolor}

\usepackage{graphicx}
\usepackage{cleveref}
\usepackage[numbers,sort&compress]{natbib}
\title{$\Sigma$-net: Ensembled Iterative Deep Neural Networks \\for Accelerated Parallel MR Image Reconstruction}
\author{%
  Jo Schlemper, Chen Qin, Kerstin Hammernik\\
  Imperial College London\\
  \texttt{\{js3611,c.qin15,k.hammernik\}@imperial.ac.uk} \\
  \And
  Jinming Duan\\
  University of Birmingham\\
  \texttt{j.duan@bham.ac.uk}
  \And
 Ronald M. Summers\\
 NIH Clinical Center \\
  \texttt{rms@nih.gov} \\
}

\begin{document}

\maketitle

\begin{abstract}
We explore an ensembled $\Sigma$-net for fast parallel MR imaging, including \emph{parallel coil networks}, which perform implicit coil weighting, and \emph{sensitivity networks}, involving explicit sensitivity maps.
The networks in $\Sigma$-net are trained in a supervised way, including content and GAN losses, and with various ways of data consistency, i.e., proximal mappings, gradient descent and variable splitting.
A semi-supervised finetuning scheme allows us to adapt to the k-space data at test time, which, however, decreases the quantitative metrics, although generating the visually most textured and sharp images.
For this challenge, we focused on robust and high SSIM scores, which we achieved by ensembling all models to a $\Sigma$-net.
\end{abstract}

\section{Introduction}
The fastMRI~\cite{zbontar2018fastmri} multicoil challenge provides a great opportunity to show how we can push the limits of acquisition speed by combining parallel magnetic resonance imaging (pMRI) and deep learning.
Recent works on pMRI show the success of learning a fixed iterative reconstruction scheme, involving the MR forward model in various ways~\cite{hammernik2018, aggarwal2017modl, duan2019vs, schlemper2019data, qin2018convolutional}.
In this work, we explore different types of reconstruction networks for pMRI: (1) parallel coil networks (PCNs), that learn implicit weighting of the single coils, and (2) sensitivity networks (SNs), that require explicit coil sensitivity maps~\cite{pruessmann1999sense,uecker2014espirit}.
We investigate different ways of incorporating data consistency (DC) and train the networks in both a supervised and semi-supervised manner.
Instead of choosing a single model for pMRI reconstruction, we increase the robustness by ensembling the individual model reconstructions, termed $\Sigma$-net.
To meet the quantitative evaluation criteria, we introduce - exclusively for this challenge - a style transfer layer (STL), which maps the contrast of SNs to root-sum-of-squares (RSS) reconstructions.
%

\section{Methods}
We explore a variety of network architectures, loss functions and learning strategies.
In the following, we give a short overview of the different architectures and show how we achieve an ensembled $\Sigma$-net.

\subsection{Learning unrolled optimization}
All of our models are based on learning a fixed iterative scheme~\cite{hammernik2018, aggarwal2017modl, duan2019vs, schlemper2019data}. In general, this has the following form to obtain a reconstruction $x$ from k-space data $y$ involving a linear forward model $A$
\begin{align*}
    x^{t+\frac{1}{2}} = x^t - f_{\theta^t}(x^t),\quad
     x^{t+1} = g(Ax^{t+\frac{1}{2}},y), \quad 0 \leq t < T.
\end{align*}
Here, $f_\theta$ represents the neural network-based reconstruction block, $g$ denotes a DC layer and $T\!\!=\!\!9$ is the number of steps. Each reconstruction block has the form of an encoding-decoding structure, such as U-net \cite{ronneberger2015u} and Down-Up CNNs~\cite{yu2019didn}. The DC is considered in various ways. We investigate DC as gradient descent (GD)~\cite{hammernik2018}, proximal mappings (PM)~\cite{schlemper2017deep,aggarwal2017modl} and variable splitting (VS)~\cite{duan2019vs}. For pMRI involving explicit sensitivity maps, the PMs are solved numerically using conjugate gradient~\cite{aggarwal2017modl}.

\subsection{Network architectures}
We investigate two types of architectures for pMRI reconstruction. \emph{Parallel coil networks} (PCNs) reconstruct individual coil images $x=[x_1,\ldots,x_Q]$ for $Q$ coils. The network $f_\theta$ is realized by a U-net~\cite{ronneberger2015u} with $Q$ complex-valued input and output channels and learns implicit coil weightings. For \emph{sensitivity networks} (SNs), the coil combination is defined in the operator $A$ using explicit coil sensitivity maps as in~\cite{hammernik2018}. To overcome field-of-view issues in the SNs, we use an extended set of two coil sensitivity maps according to~\cite{uecker2014espirit}, hence, reconstructing $x=[x_1,x_2]$. In this case, the network $f_\theta$ has two complex-valued input and output channels and is modelled by a Down-Up network~\cite{yu2019didn}. The final reconstruction $x_{\text{rec}}$ is obtained by RSS combination of the individual channels of $x^T$.

\subsection{Supervised and semi-supervised learning}
We trained individual networks for the acceleration factors  $R\!\!=\!\!4$ and  $R\!\!=\!\!8$ as well as contrasts PD and PDFS of the fastMRI~\cite{zbontar2018fastmri} training set. 
The networks were trained using $\ell_1$ and SSIM loss~\cite{zhao2016loss, hammernik2017l2} between the reference $x_{\text{ref}}$ and the reconstruction $x_\text{rec}\!\!=\!\!x^T$, involving the binary foreground mask $m$,
\begin{align*}
    \ell_\text{base}(x_\text{rec}, x_\text{ref}) = \text{SSIM}(m \odot |x_\text{rec}|, m \odot |x_\text{ref}|) + \lambda \ell_1(m \odot |x_\text{rec}|, m \odot |x_\text{ref}|)
\end{align*}
where $\odot$ is the pixel wise product and $|\cdot|$ denotes the RSS reconstruction to combine the individual output channels.
The parameter $\lambda\!\!=\!\!10^{-3}$ was chosen empirically to match the scale of the two losses.
We trained for 50 epochs using RMSProp with learning rate $10^{-4}$, reduced by 0.5 every 15$^{\text{th}}$ epoch. 

\paragraph{Least Squares GAN (LSGAN)}  
The trained model was further finetuned for 10 epochs with learning rate $5\!\!\times\!\! 10^{-5}$ using an LSGAN loss
%
$
    \ell_\text{GAN}(x_\text{rec}, x_\text{ref}) = \gamma\ell_\text{base}(m \odot |x_\text{rec}|, m \odot |x_\text{ref}|) + \ell_\text{LSGAN}(m \odot |x_\text{rec}|, m \odot |x_\text{ref}|),
$
with the same architecture as in \cite{ledig2017photo}.
The parameter $\gamma$ was chosen empirically via searching on the validation set and is set to $\gamma\!\! = \!\!0.1$ for PD and $\gamma\!\!=\!\!0.2$ for PDFS data.
\paragraph{Semi-supervised finetuning} To adapt to k-space data efficiently and overcome smooth reconstructions, we consider the problem $\min_\theta \frac{1}{2}\Vert Ax(\theta) - y \Vert_2^2 + \frac{\alpha}{2} \max\left(\text{SSIM}(|x(\theta)|, |x_\text{rec}|) - \beta), 0\right)^2$, where we use the initial network output $x_\text{rec}$ as a prior. We empirically chose $\alpha\!\!=\!\! 1$ and $\beta\!\!=\!\!0.008$ and finetune for 30 epochs on 4 slices of a patient volume simultaneously using ADAM  (learning rate $5\!\!\times\!\!10^{-5}$). The trained parameters are then used to reconstruct the whole patient volume.

\subsection{Experimental setup}
Using the described tools, we trained one PM-PCN, with the individual fully sampled coil images as $x_{\text{ref}}$, and four different SNs with different data consistency layers and losses, i.e., PM-SN, GD-SN, VS-SN, GD-SN-LSGAN. Additionally, we finetuned the GD-SN which we denote as GD-SN-FT. The reference $x_{\text{ref}}$ for the SNs was defined by the sensitivity-combined fully sampled data.

\paragraph{Style-transfer layer (STL)}
Although RSS is suitable surrogate for coil combination~\cite{roemer1990nmr}, we observed, however, that the gap between RSS and sensitivity-weighted images $(R\!\!=\!\!1)$ is relatively large for PDFS cases due to Rician bias of noise (see Fig.~\ref{fig:results}). To bridge this gap, we trained a STL based on a SN with  $n_f\!\!=\!\!32$ initial features. The STL was trained on the SSIM loss $\ell_\text{ST}(x_\text{rec}, x_\text{rss}) = \text{SSIM}(\text{STL}(|x_\text{rec}|), |x_\text{rss}|)$ for 10 epochs using RMSProp (learning rate $5\!\!\times\!\! 10^{-5}$).

\paragraph{Ensembling} To get robust quantitative scores, we use following ensemble to form the $\Sigma$-net
\begin{align*}
x_\text{rec} = m \odot (0.3\cdot x_\text{SN} + 0.2\cdot x_\text{PCN} + 0.5\cdot x_\text{SN-FT}) + (1-m) \odot \frac{x_\text{SN} + x_\text{PCN}}{2}.
\end{align*}
The reconstruction $x_\text{SN}$ contains the average over the SNs excluding the GD-SN-FT, which is denoted by $x_\text{SN-FT}$, and 
$x_\text{PCN}$ is the PM-PCN reconstruction.

%
%


\subsection{Data Processing}\textbf{}
We estimated two sets of sensitivity maps according to soft SENSE~\cite{uecker2014espirit} from 30 auto-calibration lines (ACLs) for $R\!\!=\!\!4$ and 15 ACLs for $R\!\!=\!\!8$ for the training and validation set.
For the test and challenge set, the sensitivity maps were computed from the provided ACLs.

To overcome the huge memory consumption of the proposed networks, we use a patch learning strategy~\cite{schlemper2017deep} where we extract patches of size 96 in frequency encoding (FE) direction without introducing new artifacts.  At test time, the network is applied to the full data.

To stabilize training, we generated foreground masks semi-automatically using the graph cut algorithm for 10 cases. This is followed by a self-supervised refinement step using a U-net with $n_f\!\!=\!\!32$ initial features. The background was replaced by the mean value, estimated from $100\!\!\times\!\!100$ noise patches of the undersampled RSS and scaled by the true acceleration factor, to match the RSS background level.

%

\section{Results}
\begin{figure}
    \centering
    \includegraphics[width=0.8\linewidth]{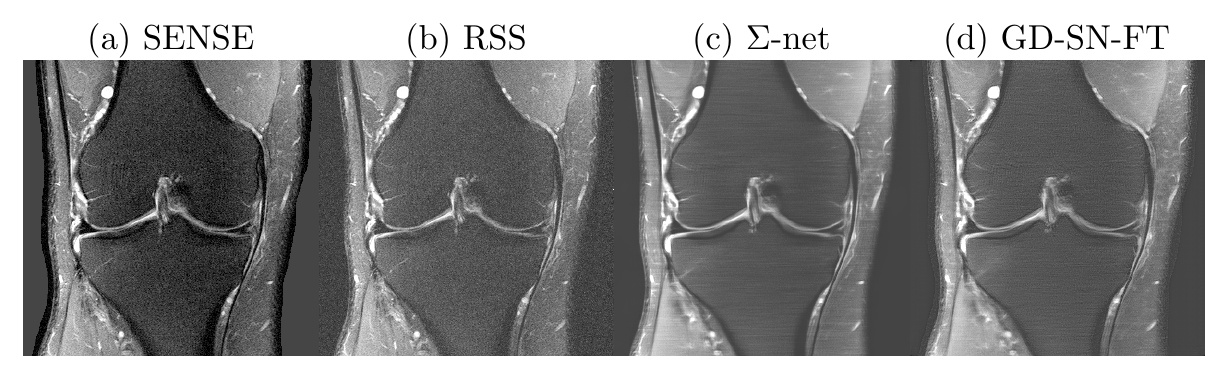}
    \caption{PDFS@1.5T (a) Fully-sampled sensitivity-weighted reference used for SN training (b) RSS target for quantitative challenge evaluation; (c) ensembled $\Sigma$-net (d) semi-supervised FT, for $R\!\!=\!\!8$.}
    \label{fig:results}
\end{figure}

We present quantitative scores on the fastMRI validation set in Tab.~\ref{tab:my_label} and qualitative results on a PDFS case for $R\!\!=\!\!8$ in~Fig.~\ref{fig:results}. The ensembled $\Sigma$-net achieves the best SSIM scores. While the scores of SN-FT are low, it appears most textured and sharp compared to the ensembled $\Sigma$-net result.

\begin{table}[h]
 \centering
 \caption{Quantitative results averaged over the whole fastMRI validation set}
 \resizebox{\linewidth}{!}{%
 \begin{tabular}{ccccccc}
 \toprule
 & \multicolumn{3}{c}{$R\!\!=\!\!4$} & \multicolumn{3}{c}{$R\!\!=\!\!8$}\\
  Method   &   NMSE & PSNR & SSIM & NMSE & PSNR & SSIM\\\midrule 
  GD-SN & 0.0069 $\pm$ 0.0243 & 38.91 $\pm$ 6.46 & 0.9136 $\pm$ 0.1320 &   0.0130 $\pm$ 0.0503 & 35.39 $\pm$ 4.68 & 0.8809 $\pm$ 0.1341\\
  PM-SN & 0.0071 $\pm$ 0.0250 & 38.81 $\pm$ 6.64 & 0.9135 $\pm$ 0.1340 &  0.0137 $\pm$ 0.0508 & 35.12 $\pm$ 4.59 & 0.8790 $\pm$ 0.1377\\
  VS-SN &  0.0069 $\pm$ 0.0265 & 38.98 $\pm$ 6.57 & 0.9138 $\pm$ 0.1326 & 0.0118 $\pm$ 0.0511 & 36.15 $\pm$ 5.19 & 0.8842 $\pm$ 0.1374\\
  GD-SN-LSGAN & 0.0069 $\pm$ 0.0267 & 38.99 $\pm$ 6.55 &  0.9137 $\pm$ 0.1322 & 0.0118 $\pm$ 0.0538 & 36.18 $\pm$ 5.18 &  0.8841 $\pm$ 0.1367\\
  GD-SN-FT & 0.0069 $\pm$ 0.0125 & 38.46 $\pm$ 6.10 & 0.9085 $\pm$ 0.1327 & 0.0107 $\pm$ 0.0124 & 36.01 $\pm$ 4.59 & 0.8808 $\pm$ 0.1352 \\
  PM-PCN & 0.0064 $\pm$ 0.0117 & 38.61 $\pm$ 5.67 & 0.9127 $\pm$ 0.1199 & 0.0115 $\pm$ 0.0150 & 35.56 $\pm$ 4.42 & 0.8785 $\pm$ 0.1277\\
  $\Sigma$-net & {\bf0.0055 $\pm$ 0.0118} & {\bf39.57 $\pm$ 6.42} & {\bf0.9205 $\pm$ 0.1234} & {\bf 0.0091 $\pm$ 0.0150} & {\bf36.83 $\pm$ 4.97} & {\bf0.8917 $\pm$ 0.1317} \\
     \bottomrule
     \end{tabular}}
\label{tab:my_label}
\end{table}

\section{Conclusion}
This work shows the results for various PCNs and SNs, which are included in an ensembled $\Sigma$-net.
We observe that the SNs perform similar and the final ensembling reduces random errors made by the individual networks.
Semi-supervised finetuning is a promising way to get the texture and noise back from the original k-space data.
Although the GD-SN-FT would suit the human eye best, these results demonstrate, however, again that quantitative metrics do not coincide with the visual perception.
The challenge of this multicoil challenge was the RSS reference, requiring a STL for SNs to match the contrast.
This has no practical relevance and visually decreases the quality of our initial results.
Hence, future work will focus on evaluating the SNs on sensitivity-combined reference images.

\subsubsection*{Acknowledgements}

The work was partially funded by EPSRC Programme Grant
(EP/P001009/1).

\medskip
\small

\bibliography{ensembling}

\end{document}